\documentclass[12pt,preprint]{aastex}

\usepackage{epsfig}

\begin{document}

\newcommand{\beq}{\begin{equation}}
\newcommand{\enq}{\end{equation}}
\newcommand{\ba}{\begin{eqnarray}}
\newcommand{\ea}{\end{eqnarray}}
\newcommand{\siml}{\lower4pt \hbox{$\buildrel < \over \sim$}}
\newcommand{\simg}{\lower4pt \hbox{$\buildrel > \over \sim$}}
\newcommand{\Mesz}{M\'esz\'aros~}
\newcommand{\epsd}{\epsilon_d}
\newcommand{\cm}{{\rm cm}}
\newcommand{\erg}{{\rm erg}}
\newcommand{\si}{{\rm s}^{-1}}
\newcommand{\msun}{{M_\odot}}
\newcommand{\rsun}{{R_\odot}}
\def\fm{\hbox{$.\!\!^{\rm m}$}}
\def\farc{\hbox{$.\!\!^{\prime\prime}$}}
\def\fdg{\hbox{$.\!\!^\circ$}}
\def\degr{\hbox{$^\circ$}}
\def\arcsec{\hbox{$^{\prime\prime}$}}
\def\ha{H$\alpha${}}
\def\hb{H$\beta${}}
\def\>{$>$}
\def\<{$<$}
\def\bsl{$\backslash$}
\def\simlt{\lower.5ex\hbox{$\; \buildrel < \over \sim \;$}}
\def\simgt{\lower.5ex\hbox{$\; \buildrel > \over \sim \;$}}
\def\ch2{$\chi^{2}$}
\def\ii{\'{\i}}
\def\aa{\'{a}}
\def\ee{\'{e}}
\def\oo{\'{o}}

\def\sT{\sigma_{\rm T}}
\def\dd{{\rm d}}

\def\be{\begin{equation}}
\def\ee{\end{equation}}
\def\curl{{\rm curl}}
\def\bA{{\,\mathbf A}}
\def\bB{{\,\mathbf B}}
\def\bE{{\,\mathbf E}}
\def\bj{{\,\mathbf j}}
\def\bv{{\,\mathbf v}}
\def\Erot{E_{\rm rot}}
\def\Ek{E_{z,\mathbf k}}
\def\Phik{\Phi_{\mathbf k}}
\def\rhok{\rho_{\mathbf k}}
\def\jk{j_{\mathbf k}}
\def\Ak{A^\prime_{z,\mathbf k}}
\def\jB{j_B}
\def\RNS{R_{\rm NS}}
\def\gres{\gamma_{\rm res}}
\def\Ecurl{E^{\rm curl}}
\def\Rmax{R_{\rm max}}
\def\Rlc{R_{\rm lc}}
\def\tT{\tau_{\rm T}}

\newbox\grsign \setbox\grsign=\hbox{$>$} \newdimen\grdimen \grdimen=\ht\grsign
\newbox\simlessbox \newbox\simgreatbox \newbox\simpropbox
\setbox\simgreatbox=\hbox{\raise.5ex\hbox{$>$}\llap
     {\lower.5ex\hbox{$\sim$}}}\ht1=\grdimen\dp1=0pt
\setbox\simlessbox=\hbox{\raise.5ex\hbox{$<$}\llap
     {\lower.5ex\hbox{$\sim$}}}\ht2=\grdimen\dp2=0pt
\setbox\simpropbox=\hbox{\raise.5ex\hbox{$\propto$}\llap
     {\lower.5ex\hbox{$\sim$}}}\ht2=\grdimen\dp2=0pt
\def\simgt{\mathrel{\copy\simgreatbox}}
\def\simlt{\mathrel{\copy\simlessbox}}
\def\simprop{\mathrel{\copy\simpropbox}}

\slugcomment{Astrophysical Journal, vol. 664, in press}
\title{Thermalization in Relativistic Outflows and the Correlation between
Spectral Hardness and Apparent Luminosity in Gamma-ray Bursts}

\author{C. Thompson}
\affil{CITA, 60 St. George St., Toronto, ON, M5S 3H8}
\author{P. \Mesz}
\affil{Department of Astronomy \& Astrophysics, PSU}
\author{M.J. Rees}
\affil{Institute of Astronomy, Cambridge, U.K.}

\begin{abstract}
 We present an interpretation of the phenomenological relations between the
 spectral peak, isotropic luminosity and duration of long gamma ray bursts that
 have been discovered by Amati et al., Ghirlanda et al., Firmani et al., and
 Liang \& Zhang. In our proposed model, a jet undergoes internal dissipation
 which prevents its bulk Lorentz factor from exceeding 1/$\theta$ ($\theta$ 
 being the jet opening angle)  until it escapes from the core of its 
 progenitor star, at a radius of order $10^{10}$ cm; dissipation may continue
 at larger radii. The dissipated radiation will be partially thermalized, 
 and we identify its thermal 
 peak (Doppler boosted by the outflow) with $E_{\rm pk}$. The radiation comes,
 in effect, from within the jet photosphere. The non-thermal, high energy 
 part of the GRB emission arises from Comptonization of this radiation by
 relativistic electrons and positrons outside the effective photosphere.
 This model can account naturally not only for the surprisingly small scatter
 in the various claimed correlations, but also for the normalization, as well
 as the slopes.  It then has further implications for the jet energy, the
 limiting jet Lorentz factor, and the relation of the energy, opening angle
 and burst duration to the mass and radius of the stellar stellar progenitor.
 The observed relation between pulse width and photon frequency can be reproduced
 by inverse-Compton emission at $\sim 10^{14}$ cm from the engine, but there 
 are significant constraints on the energy distribution and isotropy
 of the radiating particles.
\end{abstract}

\keywords{gamma rays: bursts; supernovae: general; radiation 
mechanisms: non-thermal}

\section{Introduction}
\label{sec:intro}

The mechanism by which gamma-rays are created in the transient events
know as gamma-ray bursts (GRBs) 
has remained surprisingly resistant to interpretation.
Two key ideas have guided theoretical efforts.
The first is that thermalization must be rapid and nearly complete
close to the engine.  A fireball is created, which could reach 
a temperature as high as $\sim 1$ MeV within a region the size of
a neutron star or stellar-mass black hole (Paczy\'nski 1986; Goodman 1986).  
The second idea is that strongly non-thermal particle distributions 
are expected at large distances from the central engine, e.g. when the
relativistic ejecta collide with the ambient medium (Rees \& Meszaros 1992).
Continuing dissipation within the outflow itself can also be expected,
due to the formation of internal shocks (Rees \& Meszaros 1994), or the
release of magnetic energy by reconnection (Thompson 1994; 
Drenkhahn \& Spruit 2002) or global current-driven instabilities
(Giannios \& Spruit 2006; Lyutikov \& Blandford 2003).

The prompt emission of a GRB appears at first sight to be highly non-thermal.
A high-energy cutoff is seen only in a modest fraction of GRB spectra
(Pendleton et al. 1997; Ryde 2005).  Often the low-energy spectrum
appears to be harder than would be consistent with optically thin synchrotron 
emission (Crider et al. 1997; Ghirlanda, Ghisellini, \& Cellotti 2003).
A discrete thermal bump is usually not apparent in the usual spectral
analyses, although Ryde (2005) has shown that up to 30\% of long GRB can be 
interpreted as a combination of a thermal peak plus a power law spectrum.  
Intriguing evidence has, furthermore, emerged, which indicates that the gamma-ray 
emission is seeded by radiation with a nearly black body spectral distribution.
This evidence comes from measurements of the photon
energy $E_{\rm pk}$ at which the GRB energy spectrum 
$E_\gamma^2 dN_\gamma/dE_\gamma$ has a maximum.  

There has been longstanding interest in the distribution of
$E_{\rm pk}$ values in GRBs.
BATSE measured a distribution that is clustered around
$200$ keV (e.g. Preece et al. 1999).  Observational selection might 
explain the high-energy cutoff to this distribution:  the GRB spectrum is
harder below the peak and so a lower flux is measured within the BATSE band
from bursts with large $E_{\rm pk}$ (Piran \& Narayan 1996).  Nonetheless,
the existence of a substantial population of GRBs with $E_{\rm pk}$ higher
than $\sim 1$ MeV remains an open question.  

Our knowledge of the $E_{\rm pk}$ distribution has improved dramatically
with the localizations provided by BeppoSAX and HETE-II, which have
allowed a direct measurement of the distances to bursts
(see Piran 2005; \Mesz 2006, for detailed references).
The value of $E_{\rm pk}$ in the cosmological rest frame of the burst source
correlates well the isotropic gamma-ray energy,
\be\label{eq:amati}
E_{\rm pk}^i = 100\,\left({E_{\rm iso}\over 10^{52}~{\rm erg}}\right)^{0.5}
\;\;\;\;\;{\rm  keV}
\ee
(Amati et al. 2002; Lamb et al. 2005).  Bursts with lower peak energies
are therefore substantially dimmer.  Related correlations have
been obtained by correcting $E_{\rm iso}$ for the beaming angle,
as inferred from the measurement of breaks in the afterglow
light curve (Ghirlanda et al. 2004; Liang \& Zhang 2005), and by
measuring the joint correlation of $E_{\rm pk}$ with $E_{\rm iso}$
and the burst duration (Firmani et al. 2006).

The breadth of the intrinsic $E_{\rm pk}^i$ distribution provides valuable
information about the emission physics.
In the sample of bursts with measured redshifts, $E_{\rm pk}^i$
generally lies in the range 100 keV - 1 MeV (Amati 2006).  These values
are generally
lower than what would be expected from thermalization close to the
engine.  Although a broader distribution of $E_{\rm pk}$ can result from
prompt thermalization followed by adiabatic cooling
(e.g. Thompson 1994; M\'esz\'aros \& Rees 2000, M\'esz\'aros et al. 2002), 
a more plausible explanation is that thermalization takes place at greater
distances from the engine than was suggested by the original fireball models.
The relation between $E_{\rm iso}$ and $E_{\rm pk}^i$ that results from
this type of distributed heating has been investigated recently
(Rees \& M\'esz\'aros 2005; Thompson 2006; Pe'er et al. 2006).
The peak energy can remain quite high at large distances
from the engine, after allowance is made
for beaming and bulk relativistic motion.   Continuing dissipation in 
the jet is a natural consequence of its interaction with the
core and envelope of a Wolf-Rayet star.  

Bursts with peak energies less than 50 keV (X-ray flashes) have been
detected by Ginga and HETE-II (Strohmayer et al. 1998; Sakamoto et al. 
2005), but few have measured redshifts.  Their $T_{90}$ distribution is, 
nonetheless, remarkably similar to that of the harder spectrum bursts, 
which suggests that the low $E_{\rm pk}$ is not primarily due to a large 
cosmological redshift.  Since the formation of a nearly black body 
spectrum is possible only out to some distance from
the engine, one expects that the thermal peak energy must cover only a
limited range.  In fact, as we will see, quasi-thermal models of the prompt
GRB emission over $\sim 10$ seconds cannot easily be extended below 
$E_{\rm pk}^i \sim 50$ keV without invoking strong adiabatic cooling.  
Peak energies much higher than 1 MeV also cannot easily be obtained 
without invoking upscattering by relativistic particles.

It has  been noted (Nakar \& Piran 2005) that among long bursts 
in the BATSE catalog whose redshift is unknown, there may be up to
25-30\% whose low luminosity  and hard spectra are 
incompatible with the Amati et al. relation.  The implication is that
this relation may mark out a limiting value of the spectral peak energy
at a given apparent luminosity.
For the quasi-thermal model discussed here, we point out that in fact at low
luminosities a departure from the Amati et al. 
relation is expected, which works
in the sense of the above discrepancy:  relativistic jets with large
opening angles and relatively low isotropic luminosities should produce 
spectrally harder bursts.

In addition to providing seeds for the prompt gamma-ray emission,
a thermal photon field also can play an important role in accelerating
the outflow (Pacyznski 1990; Shemi \& Piran 1990).  Phenomenological
scaling relations such as the Amati et al. relation therefore 
provide crucial information about
the {\it relative sequence} of thermalization and acceleration.
If thermalization occurs first, at some characteristic radius, then
a tight relation between $E_{\rm pk}$ and $E_{\rm iso}$ follows
naturally.  Moreover, it has been suggested that the slope of the 
$E_{\rm pk}-E_{\rm iso}$ relation
can be reproduced most easily if the internal heating is driven by non-radial 
shear instabilities near the base of the jet -- where its Lorentz factor 
$\Gamma$ is comparable to the inverse of the
opening angle $\theta$ -- as opposed to radial instabilities in a relativistic
flow with $\Gamma \gg 1/\theta$ (Thompson 2006).  This raises the 
possibility that components of the jet which emerge from a Wolf-Rayet star
with low baryon loading but also a low entropy (e.g. the jet core) 
will attain lower terminal Lorentz factors, 
and may manifest themselves in the afterglow phase.

The approach adopted in this paper is two-fold.   After an initial outline
in \S 2 of the gist of our thermal interpretation of the Amati et al. relation,
we begin by working backward from the observed correlation between spectral
peak energy and isotropic burst energy.   In \S 2
we show that thermalization must continue out to a large distance
($\sim 10^{10}$ cm) from the engine in order to reproduce the
normalization of the Amati et al. relation.  We show that thermalization
at this radius is reasonable, and derive a condition for the outflow 
to generate enough photons to establish a blackbody spectral distribution
(\S 2.1).  In particular, the magnetic field must carry at least
$\sim 10\%$ of the outflow luminosity.  The acceleration of the outflow
is addressed in \S 2.2, where it is shown that the blackbody photons
can effectively accelerate the entrained baryons (and magnetic field)
up to a limiting Lorentz factor of 100-500.  In \S 2.3 we 
examine the effect on the $E_{\rm pk}$-$E_{\rm iso}$ relation of having
thermalization occur before or after the acceleration of the jet
material.  In the second case, $E_{\rm pk}$ has a much weaker dependence
on $E_{\rm iso}$ than is observed.   

The alternative works much better.  The normalization and slope of the 
Amati et al. relation follow immediately from two simple assumptions:  
first, that the total jet energy is regulated to a value close to the net 
gravitational binding energy of the Wolf-Rayet core and, second, 
that the GRB-emitting component of the jet has $\Gamma \sim 1/\theta$ 
at the core boundary.  This simple model predicts
the joint correlation of $E_{\rm pk}$ with $E_{\rm iso}$ and the
burst duration $t_j$ derived from the data by Firmani et al. (2006).
The possibility of a range of jet energies is examined in \S 2.4,
where it is shown that the Ghirlanda et al. (2004) relation is consistent
with an approximate scaling $E_j \sim \theta_j^{-1}$.  Thermalization
can also continue out to the photosphere in jets with heavier baryon
loadings, and a minimum thermal peak energy of 50 keV (without adiabatic
cooling) is derived in \S 2.5.  The implications of the 
$E_{\rm pk}$-$E_{\rm iso}$ relation of short GRBs for our model are 
outlined in \S 2.6, where it is shown that a large thermalization radius, 
similar to that of the long GRBs ($R_0 \sim 10^{10}$ cm) is implied.  
The pulses of GRBs are generally narrower at higher frequencies (Fenimore et 
al. 1995), and it is argued in \S 3.1 that this is nicely consistent with IC
scattering of the seed thermal photons by a power-law distribution
of electrons (and positrons) at $\sim 10^{14}$ cm from the engine.  
However, this explanation for the pulse duration has an important
implication:  the IC emission by particles with relativistic energies
must be {\it beamed} in the bulk frame.  In addition, the distribution
of particle energies must have a lower cutoff at transrelativistic
energies ($\gamma_{\rm min} \sim 1$).  We consider the implications
for models of particle heating (by turbulence and shocks) in \S 3.2.
The paper conclusions with some summary observations in \S 3.3.

\section{{\bf Quasi-Thermal} Model of Burst Emission}

We assume here that the flow can be characterized by
a mean Lorentz factor $\Gamma(r)$ and opening angle $\theta(r)$ at each radius $r$.
The opening angle can vary with radius if a non-radial magnetic field 
plays a role in collimation and acceleration (e.g. Lynden-Bell 2003;
Vlahakis \& K\"onigl 2003).  An observer viewing the outflow face on
sees a luminosity
\be
L_{\rm iso}(r) = {2L_j\over\theta_j^2(r)}\;\;\;\;\;\;(\theta_j \ll 1),
\label{eq:LisoLj}
\ee
where $L_j$ is the total output of the engine and $\theta_j$ is the
half-opening angle of the outflow.

A fraction $\varepsilon_{\rm bb}$ of the outflow energy is assumed to
be completely thermalized at a radius $R_0$, where the outflow Lorentz factor is
$\Gamma_0$.   In the present context, we
will find that the temperature $T'$ in the rest frame of the outflow is low enough 
that thermally-created pairs do not contribute
significantly to the specific heat.  The bulk frame thermal energy density 
is then $\simeq aT'^4$, where $a$ is the Stefan-Boltzmann constant.  
An observer at rest with respect to the engine sees a temperature
\be\label{eq:Lphot}
T_{\rm obs} = {4\over 3}\Gamma_0 T_0' = 
\varepsilon_{\rm bb}^{1/4}\,\left({\Gamma_0\over R_0}\right)^{1/2}
\left({16L_{\rm iso}\over 27\pi ac}\right)^{1/4}.
\label{eq:LBB}
\ee
It is reasonable to assume that strong thermalization takes place only at
radii comparable or less than the radius of the stellar progenitor, 
$R_0 \la R_\ast$,
e.g. that being the limiting radius where oblique shocks can cause dissipation.
At this radius, one would expect the bulk Lorentz factor $\Gamma_0 (R_0) \sim 
1/2\sqrt{3}\theta_j$ where $\theta_j$ is the jet opening angle at the surface, 
since above this Lorentz factor shear-driven modes in the jet
will have no time to
grow on a lengthscale $\sim R_\ast\theta$ (in the bulk frame). Thus, using
equations (\ref{eq:LisoLj}) and (\ref{eq:Lphot}), and assuming 
$\Gamma (R) \sim \theta_j^{-1}$ and $L_j = E_j/t_j\sim$ constant
(as motivated by Frail et al. 2001, who find $E_j \sim$ const) one obtains 
\be
E_p \propto R_0^{-1/2} t_j^{-1/4} E_{\rm iso}^{1/2}~
\label{eq:amati-theo}
\ee
(Thompson 2006).
This is essentially the relation discovered by Amati et al. (2002), if we consider
that $R_0\sim R_\ast \sim$ constant (e.g., all progenitors of long GRBs are
lacking hydrogen envelopes) and the dispersion in $t_j^{1/4}$ ($10^{1/4}$ for
the majority of long GRBs) is smaller than that in $E_{iso}^{1/2}$ (approximately $10^{4/2}$, two decades).

This simple derivation of the Amati et al. 
relation involves making some reasonable
physical assumptions. However, it is more illuminating to work backwards, and
use the observed relation (\ref{eq:amati}) to derive implications
for the physics of the model and the progenitor. Combining the black body law 
(\ref{eq:LBB}) with eq. (\ref{eq:amati}) gives\footnote{Throughout,
we use the shorthand $X = X_n\times 10^n$, 
where the quantity $X$ is measured in c.g.s. units.}
\be\label{eq:rgamat}
{R_0\over\Gamma_0} = 6\times 10^9\,\varepsilon_{\rm bb}^{1/2}
E_{\rm iso\,52}^{-1/2}\,t_{j\,1}^{-1/2}\;\;\;\;\;\;{\rm cm}.
\ee
One observes that the Amati et al. relation can be reproduced, but only if 
thermalization takes place at a radius $R_0$ that is much larger than 
the engine radius (in the conventional picture where the engine
is an accreting black hole or possibly a rapidly-spinning magnetar).
If the motion of the jet is only mildly relativistic at this point
then $R_0$ is comparable to the core radius of a Wolf-Rayet star.

\subsection{Thermalization in a Plasma of Moderate Scattering Optical Depth}
\label{s:therm}

Consider now the consequences of heating an ionized plasma that is 
optically thick to scattering, but optically thin to free-free absorption.   
Under what circumstances does the radiation field relax to a nearly
black body distribution?  To get a sense of the Thomson optical
depth $\tau_T$ that is required, we first assume that the radiation spectrum
is nearly black body, and then examine the self-consistency of this
assumption.

Synchrotron emission by relativistic electrons (and positrons) is
generally a copious source of soft photons.  In spite of this, a 
minimal magnetic field is required for synchrotron emission to supply
enough photons to establish a black body gas.  We suppose that
a fraction $\varepsilon_{\rm sa}$ of the energy density $U'$ of the outflow
is transferred to electrons that radiate at the synchrotron 
self-absorption frequency (at a Lorentz factor $\gamma_{\rm sa}$).  
A fraction $\varepsilon_B/(\varepsilon_B + \varepsilon_\gamma)$ 
of the energy of these relativistic electrons
is released to synchrotron photons (as opposed to IC scattering of the 
ambient radiation field, with a total energy density $U'_\gamma
=\varepsilon_\gamma U'$ from all sources).  
The number of (new) synchrotron photons is
\be\label{eq:nsynch}
n'_{\gamma\,\rm synch} \sim 
   \left({\varepsilon_B\over \varepsilon_B + \varepsilon_\gamma}\right)
   {\varepsilon_{\rm sa} U'\over 0.3\gamma_{\rm sa}^2 \hbar eB'/m_ec}.
\ee
Let us compare this expression with the number density of photons 
in a black body gas (of temperature $T_{\rm bb}'$) that carries
a fraction $\varepsilon_{\rm bb}$ of the outflow energy.
The characteristic self-absorption frequency can be obtained from
\be\label{eq:gamsat}
\gamma_{\rm sa}^5 \simeq {\tau_{\rm T\, sa}\over \alpha_{\rm em}}
\,\left({B'\over B_{\rm QED}}\right)^{-1},
\ee
where $\tau_{\rm T\,sa} = \sigma_T n_e'(\gamma_{\rm sa}) ct'$ 
is the Thomson depth through the 
electrons of energy $\gamma_{\rm sa}\,m_ec^2$.  The net output in
synchrotron photons at the self-absorption frequency is given by
\be
\varepsilon_{\rm sa} U'  \sim  n_e'(\gamma_{\rm sa})\times
{4\over 3}\gamma_{\rm sa}^2 \sigma_T {B'^2\over 8\pi} ct'. 
\ee
The optical depth $\tau_{\rm T\,sa}$ is therefore regulated to a value 
\be
{4\over 3}\gamma_{\rm sa}^2\tau_{\rm T\,sa} \sim 
{\varepsilon_{\rm sa}\over\varepsilon_B}.
\ee
Combining this expression with eq. (\ref{eq:gamsat}) gives
\be\label{eq:gamsatb}
\gamma_{\rm sa} \simeq  
\left({\varepsilon_{\rm sa}\over \varepsilon_B \alpha_{\rm em}}\right)^{1/7}
\,\left({B'\over B_{\rm QED}}\right)^{-1/7}.
\ee
The relative numbers of synchrotron and blackbody photons are obtained
by substituting eq. (\ref{eq:gamsatb}) and the 
relation $(k_{\rm B}T_{\rm bb}'/m_ec^2)(B'/B_{\rm QED})^{-1/2}
= 1.7\,(\varepsilon_{\rm bb}/\varepsilon_B)^{1/4}$
into eq. (\ref{eq:nsynch}),
\be\label{eq:nratio}
{n'_{\gamma\,\rm synch}\over n'_{\rm bb}(T'_{\rm bb})}
\sim 10{\varepsilon_{\rm sa}^{5/7}\,\varepsilon_B^{13/14}
\over \varepsilon_{\rm bb}^{9/14}\,(\varepsilon_\gamma + \varepsilon_B)}\,
\left({k_{\rm B}T'_{\rm bb}\over m_ec^2}\right)^{-3/7}.
\ee
Recall that $\varepsilon_{\rm sa}$ labels the net energy that is
injected into relativistic particles of energy $\gamma_{\rm sa}m_ec^2$.
For example, if the energy injected into relativistic electrons
is distributed uniformly over Lorentz factor and comprises a fraction
$\varepsilon_{\rm nth}$ of the total, then $\varepsilon_{\rm sa}
= \varepsilon_{\rm nth}/\ln(\gamma_{\rm max}/\gamma_{\rm min}) \sim
0.1\,\varepsilon_{\rm nth}$.   The right-hand side of 
eq. (\ref{eq:nratio}) can be larger than unity, but only if
\be
{\varepsilon_B\over \varepsilon_{\rm bb}}
\ga \left({\varepsilon_{\rm nth}\over \varepsilon_{\rm bb}}\right)^{-10/13}
\left({k_BT_{\rm bb}'\over m_ec^2}\right)^{6/13}.
\ee
We conclude that photon creation in the
outflow can be rapid enough to create a black body photon gas, but
only if the magnetic field carries more than $\sim 10$ percent of the
total dissipated energy.

Soft photons injected into the outflow are
redistributed in frequency by multiple Compton
scatterings.  When the electrons and the photons have the same
temperature, there is no net transfer of energy from electrons to photons,
but soft photons will increase their energy, at a rate
$\dot E_\gamma/E_\gamma \simeq  4(k_{\rm B}T_e'/m_ec^2)\sigma_T n_e' c$.
The efficient redistribution of photons in frequency via Compton scattering
requires that 
\be\label{eq:ycompmin}
\left({k_{\rm B}T_e'\over m_ec^2}\right)\,\tau_T \ga 1,
\ee
where $\tau_T$ is the Thomson depth through the warm electrons with
a bulk-frame temperature $T_e'$.  From eq. (\ref{eq:ycompmin}), one 
obtains a (conservative) lower bound on the Thomson depth that guarantees
efficient thermalization of the soft seed photons at an effective
temperature $T_{\rm bb}'$,
\be\label{eq:tauTmin}
\tau_T \ga \left({k_{\rm B}T_{\rm bb}'\over m_ec^2}\right)^{-1}.
\ee
We will adopt this value of the optical depth when evaluating
the temperature of the radiation that emerges from outflows with
a relatively high baryon loading.

At energies above the thermal
peak, a power component can arise in the usual manner from shocks and 
synchrotron/IC radiation above the photosphere, and/or a comptonized tail
in the photosphere itself.  This approximation of the spectral peak determined
by a black-body temperature $E_{\rm pk} \sim 3k_{\rm B}T$ is only slightly changed if one 
considers the Compton equilibrium in dissipating photosphere which is 
optically thick to scattering, $E_{\rm pk}\simeq 3k_{\rm B}T (1+A^{-1})$, where $A \sim 1$ 
is the photon to electron energy density ratio (Pe'er et al. 2005).  

\subsection{Adiabatic Cooling and Radiative Acceleration of the Outflow}

This fireball radiation may be cooled adiabatically if it is
generated close enough to the engine, so that the fireball Lorentz factor
saturates at a value $\eta = L_j/\dot M_jc^2$ before the photons and
the electrons are decoupled from each other.   A condition for the neglect
of adiabatic cooling is obtained by comparing the saturation radius $R_{\rm sat}$ with
the photospheric radius.   For the moment, we do not specify how
outflow is accelerated and take 
\be
\Gamma(r) = \left({r\over R_0}\right)^\alpha\,\Gamma_0\;\;\;\;\;\;
(R_\tau > r > R_0).
\ee
The saturation radius is then
\be\label{eq:rsat}
R_{\rm sat} = \left({\eta\over\Gamma_0}\right)^{1/\alpha}\,R_0.
\ee
If the electron-ion photosphere sits outside $R_{\rm sat}$, then its position is determined
by setting
\be\label{eq:phot}
\sigma_T {R_\tau\over 2\Gamma^2}
\,\left({L_{\rm iso}\over 4\pi \eta \mu c^3 R_\tau^2}\right) = 1.
\ee
Here $\mu$ is the mean mass per scattering charge.  It is useful to defining the compactness
\be
\ell_0 = {\sigma_T L_{\rm iso} \over 8\pi m_e c^3 \Gamma_0^3 R_0}
\ee
at the thermalization radius, and the `reduced' compactness 
\be
\widetilde\ell_0 = {m_e\over \mu}\ell_0.
\ee
Combining eqs. [\ref{eq:rsat}] and [\ref{eq:phot}] gives 
\be\label{eq:rratio}
{R_{\rm sat}\over R_\tau} = \left({\eta\over \Gamma_0}\right)^{(1+3\alpha)/(1+2\alpha)}\,
\widetilde\ell_0^{-1/(1+2\alpha)}.
\ee

Adiabatic cooling can be neglected if $R_{\rm sat} > R_\tau$, which corresponds to
\be
\eta > \eta_{\rm cool} = 
\widetilde\ell_0^{\alpha/(1+3\alpha)}\,\Gamma_0,
\ee
or equivalently to 
\be
{R_0\over\Gamma_0^{1/\alpha}} > {L_{\rm iso}\sigma_T\over 8\pi \mu c^3 
        \eta^{(1+3\alpha)/\alpha}}.
\ee
This expression simplifies to
\be
{R_0\over\Gamma_0} > 6\times 10^9\,{E_{\rm iso\,52}\over t_{j\,1}\eta_2^4}
\;\;\;\;\;\;{\rm cm}
\ee
in the case of a ballistically expanding outflow ($\alpha = 1$).
Comparing this expression with eq. (\ref{eq:rgamat}), one sees that
adiabatic cooling can be neglect beyond the radius where the thermal
peak energy is established, as long as $\eta = L_j/\dot M_jc^2 \ga 10^2$.

There is a close correspondence between the condition that adiabatic
cooling be absent, and the condition that the thermal photon flux be
strong enough to push the entrained baryons to a high Lorentz factor
the electron-ion photosphere (see Grimsrud \&
Wasserman 1998).   One requires a mechanism for
creating very high entropies in order to reach a terminal Lorentz factor
as high as $\sim 50$-100.  The photon field
is quasi-isotropic in a frame that moves with a Lorentz factor
\be
\Gamma_\gamma(r) = \left({r\over R_\tau}\right)\,\Gamma(R_\tau),
\ee
where 
\be
\Gamma(R_\tau) = \eta_{\rm cool}\,
  \left({\eta\over\eta_{\rm cool}}\right)^{-\alpha/(1+2\alpha)}
\;\;\;\;\;\;(\eta > \eta_{\rm cool})
\ee
is the combined Lorentz factor of the matter
and photons at the photosphere.  At $r>R_\tau$, the radiation field has
an anisotropic component in the rest frame of the matter, and the
resulting force keeps $\Gamma \simeq \Gamma_\gamma$ out to a radius where
\be
\sigma_T {r\over 2\Gamma}\,\left({L_{\rm iso}\over 
4\pi \Gamma^2\mu c^3 r^2}\right) \sim 1.
\ee
The matter therefore reaches a maximum Lorentz factor
\be
\Gamma_{\rm max} = \widetilde\ell^{1/4}(R_\tau)\,\Gamma(R_\tau),
\ee
which is 
\be
{\Gamma_{\rm max}\over \eta_{\rm cool}} = 
    \left({\eta\over\eta_{\rm cool}}\right)^{(1-\alpha)/4(1+2\alpha)}
\;\;\;\;\;\;(\eta > \eta_{\rm cool}).
\ee
When $\alpha = 1$, one recovers 
\be\label{eq:gamax}
\Gamma_{\rm max} = \widetilde\ell_0^{1/4}\Gamma_0 =
90\,L_{\rm iso\,52}^{1/4}\Gamma_0^{1/4}\,
\left({R_0\over 10^{10}~{\rm cm}}\right)^{-1/4}
\left({\mu\over m_p}\right)^{-1/4}.
\ee
The terminal Lorentz factor is smaller if the thermalization and
acceleration are offset from the engine by a distance $R_0 \gg 10^6$ cm.
When $\alpha < 1$, the terminal Lorentz factor is generally smaller
than (\ref{eq:gamax}), and saturates this bound only if $\eta$ is large
enough that the photosphere has contracted to $R_\tau \sim R_0$.  Note
also that the entrained magnetic field does not limit acceleration
by the thermal photons as long as $\Gamma_{\rm max} \gg 
(B^2/8\pi\rho c^2)^{1/3}$ (Thompson 2006).  Here $\rho$ is the 
baryon density as measured in the rest frame of the central engine.

This estimate of the terminal Lorentz factor allows for 
pair creation in the outflow outside the baryonic photosphere.  The
conversion of a modest fraction of the outflow energy to pairs would
significant decrease the inertia per scattering charge $\mu$, and therefore
increase the terminal Lorentz factor.  In the absence of pair creation,
the compactness at the electron-ion scattering photosphere is
$\ell \sim (m_p/m_e)(\eta/\Gamma)$, which is easily large enough to
allow effective collisions between photons of an energy $\sim m_ec^2$ 
in the bulk frame.  Only a tiny fraction $\sim 3/\ell$
of the outflow energy must be converted to such photons to increase
the number of light charges in the outflow.  One can expect the
relativistic ejecta to reach a terminal Lorentz factor as high as
$\sim 500$ due to this effect.

\subsection{Implications of and Models for the Amati and Firmani Relations}\label{sec:23}

One can consider two basic solutions to eq. (\ref{eq:rgamat}) for
the thermalization radius $R_0$, as deduced from the Amati et al. relation.
In the first, the acceleration of the outflow is delayed, so that
$\Gamma_0 \sim 1/\theta_0 \sim$ few at a radius $R_0$.  In the second
case, the outflow has already attained a substantial fraction of its
terminal speed, $\Gamma \sim \eta$, when thermalization occurs.  

\noindent
{\it Case 1.} When acceleration occurs after thermalization, one finds
\be\label{eq:r0int}
R_0 = 1\times 10^{10}\,(\Gamma_0\theta_j)\,{\varepsilon_{\rm bb}^{1/2}\over
E_{j\,51}^{1/2}\,t_{j\,1}^{1/2}}\;\;\;\;\;\;{\rm cm}.
\ee
Here we have rewritten eq. (\ref{eq:rgamat}) by substituting
$E_{\rm iso} = 2E_j/\theta_j^2$. 

A characteristic value of the thermalization radius follows from
the following considerations.    A certain fraction of the jet energy
is mixed with the material of the Wolf-Rayet star, thereby forming
a cocoon structure.  When the energy in this cocoon becomes comparable
to the binding energy of the Wolf-Rayet star, the star will explode
and the rate of mass accretion onto the engine (the central black hole)
must slow dramatically.  

After the jet head has reached the edge of the Wolf-Rayet core,
relativistic material will continue to flow through the created opening.
The Lorentz factor of this material generally increases away from the engine,
but at a much slower rate than the Bernoulli rate for adiabatic flow.
(This effect is observed in the simulations of an MHD jet propagating
through a stellar envelope by McKinney 2006.)
Two relativistic components of the jet can be distinguished:
a central core which moves with a Lorentz factor
$\Gamma > 1/\theta$; and an annular region sandwiched between the core and
cocoon which is susceptible to strong non-radial shear instabilities.
(These instabilities can involve several possible processes:
oblique shocks, turbulent shear viscosity, and magnetic reconnection,
the details of which are not addressed here.)

This last jet component moves with a Lorentz factor $\Gamma \sim 1/\theta$;
it is much hotter than the core but entrains only a small mass in baryons.
Its width will be equated to the distance that a signal can
propagate in the bulk frame, over the radial flow time $R_0/2\Gamma_0 c$.
This gives $\theta_0\Gamma_0 = 1/2\sqrt{3}$, where we have taken the
signal speed to be the sound speed $c/\sqrt{3}$ in an isotropic
relativistic fluid.   Substituting this expression and $E_j \simeq 10^{51}$ 
ergs (the binding energy of the core of a Wolf-Rayet star of initial mass 
$\sim 25\,M_\odot$; Woosley \& Weaver 1995) into eq. (\ref{eq:r0int})
gives
\be\label{eq:r0val}
R_0 = 3\times 10^9\,{\varepsilon_{\rm bb}^{1/2}\over
E_{j\,51}^{1/2}\,t_{j\,1}^{1/2}}\;\;\;\;\;\;{\rm cm}.
\ee
The peak energy scales with other quantities as
\be\label{eq:propt}
E_{\rm pk} \propto {E_{\rm iso}^{1/2}\over E_j^{1/4} t_j^{1/4} R_0^{1/2}}.
\ee
(Thompson 2006).

This interpretation of the Amati et al. relation has several
interesting consequences:
\begin{enumerate}

\item The thermalization radius (\ref{eq:r0val}) corresponds
closely to the radius of the CO core of a Wolf-Rayet star if
the total jet energy is set equal to the core binding energy. 
The binding energy just before the collapse is approximately 
$1\times 10^{51}(M_{\rm ZAMS}/20~M_\odot)$ in a star with 
a zero-age main sequence mass of 20-30$\,M_\odot$
(Woosley \& Weaver 1995).  
The CO core binding energy and radius depend relatively weakly on 
details of the evolution of the progenitor -- e.g. binary interaction, 
and the loss of a hydrogen or helium envelope.  The typical duration of 
a long GRB is comparable to the collapse time of the core, whereas the 
light travel time across the core is only $\sim 0.1$~s.

\item
The temperature of the thermal radiation
that is seen outside the photosphere is also insensitive to the baryon
loading, as long as $\eta$ lies below a critical value where
$\tau_T \ga (k_{\rm B}T_\gamma'/m_ec^2)^{-1}$ at the thermalization radius.
The electron-ion photosphere also lies well beyond the thermalization radius,
\be\label{eq:rratiob}
{R_\tau\over R_0} = 80\,{L_{\rm iso\,51}^{2/3}\over\eta_2^{1/3}}
\,\left({R_0\over 10^{10}~{\rm cm}}\right)^{-1/3}\;\;\;\;\;\;
(L_{\rm iso} \ga 10^{51}~{\rm ergs~s^{-1}}),
\ee
as long as the matter loading is high enough that the assumption of
complete thermalization is justified.  

\item
The relation between temperature and
isotropic energy is insensitive to continuing dissipation
over some distance outside the thermalization radius,
if the outflow Lorentz factor grows linearly with radius ($\alpha = 1$)
\be
\Gamma(r) = \Gamma_0\left({r\over R_0}\right).
\ee
The scattering
depth through the outflow decreases rapidly with radius,
$\tau_T \propto r^{-1}\gamma^{-2} \propto r^{-3}$, so that the surface
$\tau_T = (k_{\rm B}T_{\rm bb}'/m_ec^2)^{-1}$ sits at a radius
$\sim 0.2-0.4\,R_\tau$.  In other words, dissipation can continue
over a factor $20-30$ in radius beyond $R_0$ and still result in
a nearly black body spectrum;  but continuing dissipation between
this point and the photosphere will result in a broader
`greybody' spectrum (Pe'er et al. 2006).

\item
The thermal photons effectively accelerate the outflow
even outside the electron-ion photosphere.  As a result, portions of the jet
that acquire higher entropies inside the Wolf-Rayet envelope
may reach higher terminal Lorentz factors.  A 
jet may contain a cooler core that flows
with a Lorentz factor $\Gamma > 1/2\sqrt{3}\theta$ and
is subject only to weak shear-driven instabilities.  This feedback
between heating and acceleration provides an explanation for why
a thermal radiation field should generically be present in the parts of the
outflow that produce the prompt GRB emission.

\end{enumerate}

\noindent
{\it Case 2.} Now let us examine the other case, where acceleration occurs
before thermalization.  The temperature is now more sensitive to continuing
dissipation in the outflow, and the relation between $E_{\rm pk}$ and
$E_{\rm iso}$ is softer than implied by either the Amati et al. or 
Ghirlanda et al. relations.  
Suppose that the outflow Lorentz factor has saturated
at $\Gamma \simeq \eta$.  
The thermal peak energy that results from dissipating a fraction
$\varepsilon_{\rm bb}$ of the outflow energy at an optical depth
$\tau_T \simeq (k_{\rm B}T_\gamma'/m_ec^2)^{-1}$ is easily found to be
\be\label{eq:epklow}
E_{\rm pk}^i = 2.7\left({4\over 3}\eta\right)k_B T' = 
0.5\,\varepsilon_{\rm bb}^{1/6}\,{\eta^{5/3}\over L_{\rm iso\,51}^{1/6}}\,
\left({\mu\over m_p}\right)^{1/3}
\;\;\;\;\;\;{\rm keV}.
\ee
In this case, the peak energy depends strongly on the baryon loading.
In some models of the engine, e.g. those in which the outflow is
driven by a magnetic field threading the horizon of a black hole
(Blandford \& Znajek 1977; M\'esz\'aros \& Rees 1997), 
it is not obvious why $\eta$ should correlate strongly with $L_{\rm iso}$.

Recently, Firmani et al. (2006) have pointed out that a rather tight correlation 
exists between three observed quantities that are derived from measurements
of the prompt emission, namely $L_{\rm iso}$, $E_{\rm pk}$ and $t_{0.45}$, namely
$L_{\rm iso}\propto E_{\rm pk}^{1.62} t_{0.45}^{-0.49}$. 
Here $t_{0.45}$ is essentially a measure of the duration of the
prompt emission above a certain level (originally used for measuring the 
variability or spikiness of the prompt emission, by Fenimore \& Ramirez-Ruiz 
2000, and Reichart et al. 2001). Assuming that $t_{0.45}\sim t_j$, and that 
the energy output rate is $L_{\rm iso}\simeq E_{\rm iso}/t_j$, the 
Firmani relation can be written as $E_{\rm iso}\propto 
E_{\rm pk}^{5/3} t_j^{1/2}$.   The assumptions of a characteristic jet
energy and thermalization radius, which were previously used to
motivate the Amati et al. relation, imply
\beq
E_{\rm iso}\propto E_{\rm pk}^2 t_j^{1/2} R_0^{1/2}
\label{eq:thompson}
\enq
(see eq. [\ref{eq:propt}]).  
This essentially reproduces the phenomenological Firmani et al. (2006) relation.

\subsection{Correlation of $E_{\rm pk}$ with Jet Opening Angle: Ghirlanda Relation}\label{sec:24}

Ghirlanda et al. (2004) derived a relation connecting
the observed spectral peak energy $E_{\rm pk}$ to the collimation-corrected
jet energy $E_j$. The observed quantities are $E_{\rm pk}$, $E_{\rm iso}$ 
and the break time $t_b$ of the afterglow light curve.  They are related by
\beq
E_{\rm pk} \propto E_{\rm iso}^{1/2} t_{b}^{1/2} ~,
\label{eq:lz}
\enq
(Liang \& Zhang 2006, henceforth LZ; also Nava et al. 2006).

A narrower distribution of burst energies results from the jet collimation
correction in Ghirlanda et al. (2004),
but a spread of about one order of magnitude in energy remains. 
In this section, we relax
our assumption of a single burst energy, and consider the implications of
the Ghirlanda et al. relation for the jet properties within the Wolf-Rayet
envelope.

From $t_b$ and $E_{\rm iso}$ one can obtain the opening angle $\theta_j$ 
of the jet at the time of the afterglow break.  We assume that the ambient
medium has a power-law density profile, 
\beq
\rho_{ext} \propto r^{-k}.
\enq
The radius at which the break occurs is 
\beq
r(t_{b}) \propto [E_{\rm iso}/\Gamma^2(t_b)]^{1/(3-k)} 
          \propto E_j^{1/(3-k)}~,
\enq
where $\Gamma(t_{b}) \sim 1/\theta_j$ and
$E_j \sim (\theta^2/2) E_{\rm iso}$ were used. Substituting
\beq
t_{b} \sim r(t_{b})/ c \Gamma^2(t_{b})
\enq
into the observed LZ relation $E_{\rm pk}(E_{\rm iso},t_{b})$ 
[equation ((\ref{eq:lz})], one finds that $\theta_j$ cancels out and
\beq\label{eq:epkej}
E_{\rm pk} \propto E_j^{(4-k)/(6-2k)} \propto \cases{ E_j^{2/3} & for $k=0$ \cr
                                               E_j       & for $k=2$. \cr}
\label{eq:EpEj}
\enq
For a uniform external medium, $k=0$, this yields the original Ghirlanda et al.
relation $E_{\rm pk}\propto E_j ^{2/3}$, whereas for a wind, $k=2$ it yields 
$E_{\rm pk} \propto E_j$ (a form which, as pointed out by Nava et al. (2006),
is Lorentz invariant, since both quantities depend in the same manner on the 
bulk Lorentz factor).

Taking the above correlation at face value, one can work backward and
deduce a relation between the total jet energy and the opening angle of
the jet.  This relation is not unique, in that it depends on the 
index $k$ of the density profile in the medium outside the progenitor.
We focus, as before, on the interaction of the jet with the
core of the Wolf-Rayet star, and fix the thermalization radius at
the core radius $R_0$.  Combining relation (\ref{eq:epkej}) with 
the blackbody relation $E_{\rm pk} \propto [\Gamma(R_0)/R_0]^{1/2}
(E_{\rm iso}/t_j)^{1/4}$ and assuming the jet to be heated by
shear-driven instabilities ($\Gamma(R_0) \sim \theta^{-1}$), one finds
\be
E_j \propto (\theta^2\,t_j^{1/2}\,R_0)^{-2(3-k)/(5-k)}
\ee
This gives $E_j \propto \theta^{-12/5}$ for $k = 0$ and
$E_j \propto \theta^{-4/3}$ for $k = 2$.  The same scaling would, of course,
hold if we chose some other fixed radius for the formation of the 
thermal spectrum.

The Amati et al.  and Firmani et al. relations
both involve quantities that refer only to the prompt $\gamma$-ray emission.
The Ghirlanda et al. relation combines quantities 
referring to both the prompt emission and the afterglow, which introduces
an additional layer of model dependence.  For example, the jet opening angle
that is derived from the jet break time is assumed to be
the same as the opening angle in the prompt emission phase.  
A narrow range of outflow energies that is
assumed in the derivation of the Amati et al. relation in 
\S \ref{sec:23}, whereas jet break correction suggests a somewhat
broader range of energies (a factor $\sim 10$ for a wind medium).
The true range of jet energies is as yet undetermined.

\subsection{Heavier Baryon Loading and Possible Relation to 
X-ray Flashes}

Heavy baryon loading has two effects on an outflow:  it tends
to reduce the temperature of the thermal photons emerging from
the flow; and it can also lengthen the duration of the outflow
and hence of the photon signal.  The first effect is generally
encountered at lower baryon loadings than the second.  

X-ray flashes and GRBs are observed to have a similar distribution of
durations (Sakamoto et al. 2005).  
Although the X-ray flashes are fainter and so could
be observed from a smaller cosmological redshift, their intrinsic
durations cannot differ by more than a factor 2-3 from those of
the GRBs.  One thereby obtains a strong constraint on the the 
baryon loading in the outflows that emit X-ray flashes.

Consider an outflow in which the Lorentz factor has saturated 
at a value $\Gamma \simeq \eta$, and most of the energy is
in the kinetic energy of the baryons.   
The photon signal is lengthened considerably if $\eta$ exceeds
the critical value where
\be
t_j \simeq {R_\tau\over 2\eta^2 c},
\ee
namely
\be\label{eq:etadt}
\eta_{t_j} = 
\left({\sigma_T\,E_{\rm iso}\over 8\pi \mu c^4 t_j^2}\right)^{1/5} 
= 20\,E_{\rm iso\,52}^{1/5}t_{j\,1}^{-2/5}\,
\left({\mu\over m_p}\right)^{-1/5}.
\ee
The photosphere generally sits outside the saturation radius 
(\ref{eq:rsat}) when $\eta = \eta_{\rm t_j}$.  Substituting
eq. (\ref{eq:etadt}) into eq. (\ref{eq:rratio}) and taking
$\Gamma = r/R_0$ gives
\be
{R_\tau\over R_{\rm sat}} = 10\,
{E_{\rm iso\,52}^{1/15}\,t_{j\,1}^{1/5}\over\,R_{0,10}^{1/3}}
\left({\mu\over m_p}\right)^{-1/15}.
\ee
This means that dissipation must continue beyond the saturation radius
if a significant thermal photon energy flux is to be observed.  
The peak energy resulting from the dissipation of a fraction
$\varepsilon_{\rm bb} = {1\over 3}$ of the energy flux at a scattering depth 
$\tau_T \sim (k_{\rm B}T'/m_ec^2)^{-1}$ is obtained by substituting
eq. (\ref{eq:etadt}) into eq. (\ref{eq:epklow}),
\be
E_{\rm pk}^i = 2.7\left({4\over 3}\eta\right)k_B T' = 
50\,E_{\rm iso\,52}^{1/6}t_{j\,1}^{-1/2}\;\;\;\;\;\;{\rm keV}.
\ee
Note that the dependence of $E_{\rm pk}$ on $E_{\rm iso}$ in
this expression is weaker than in a blackbody.   We conclude that
values of $E_{\rm pk}^i$ as low as $\sim 50$ keV are consistent with
burst durations of $\sim 10$ s.

Lower thermal photon energy fluxes and temperatures are possible
if the dissipation occurs deeper in the outflow,
and the thermal photons are adiabatically cooled.  
A simple relation between $E_{\rm pk}$ and $E_{\rm iso}$
is not expected in this regime, due to the sensitivity of the
output spectrum to the location of the dissipation and the 
level of baryon loading.  It should be emphasized that the 
adiabatically softened thermal photons can still act as the dominant coolant
for relativistic electrons farther out in the outflow -- even if their
energy density is lower than that of the ambient magnetic field --
in some models of electron heating (e.g. Thompson 2006).  
In this case, the thermal photons must be upscattered 
in frequency to absorb a significant portion of the outflow energy,
and the effects of adiabatic cooling will be reversed.

\subsection{Implications for Long, Dim, Hard Outliers to the Amati Relation}
\label{sec:NP}

As pointed out by Nakar \& Piran (2005), a significant number of long, dim
and hard BATSE bursts fall, for any value of their unknown redshift, above 
and to the left of the nominal $E_{\rm pk}$ vs. $E_{\rm iso}$ Amati et al. relation.
A relevant point of the present thermal model (Thompson 2006) is that the 
predicted $E_{\rm pk}$-$E_{\rm iso}$ relation must transition from the nominal
$E_{\rm pk} \propto E_{\rm iso}^{1/2}$ law to a flatter black-body type law
$E_{\rm pk} \propto E_{\rm iso}^{1/4}$ at low values of $E_{\rm iso}$, when 
$\theta_j$ gets larger than a critical value (corresponding to $\Gamma(R_0) 
\sim 1/2\sqrt{3} \theta \sim  1$, or $\theta \sim 20$ degrees (from the 
Frail et al. (2001) relation $L_{\rm iso}t_j (\theta^2/2)= 5\times 10^{50}$ ergs 
which leads to $\Gamma(R_\ast)=1/2\sqrt{3}\theta = 0.9~L_{51}^{1/2} t_{j1}^{1/2}$).
In fact, the redshift-localized bursts appear to cut off, in the Nakar and
Piran (2005) simulations, at about the point where one would expect to see the 
transition, at $E_{\rm iso}\sim 10^{52}$ ergs.  

However, there could be other factors complicating the $E_{\rm pk}$-$E_{\rm iso}$
relation below $10^{52}$ ergs.  There could be a mix of different types of 
events, such off-axis, dirty fireballs that are adiabatically cooled, and 
off-center fireballs in which the reheating of soft bremsstrahlung photons is 
limited by pair annihilation and the saturation of the temperature at 30-50 
keV (SN 1998bw type events). Thus, one should be cautious on interpreting the
$E_{\rm pk}$ values of the couple of very soft HETE-II bursts that are 
frequently plotted on the Amati et al. curve.

\subsection{Low-energy Bursts Similar to GRB980425/SN1998bw}
\label{sec:grbsn}

Long bursts associated with detected supernovae (GRB/SNe) are a sub-class 
whose conformity (or not) to the Amati et al. relation is complicated. Some of these 
events 
satisfy this relation (e.g. GRB060218/SN2006aj,  Campana et al. 2006) 
while others do not (e.g. GRB980425/SN1998bw and GRB031203/SN2003lw).
The gamma-ray output of many GRB/SNe is intrinsically low, and there is
evidence pointing towards the presence of a transrelativistic ($\Gamma\sim 1$)
outflow, e.g. GRB060218/SN2006aj, GRB980425/SN1998bw, GRB031203/SN2003lw. 
Other GRB/SNe which are not particularly dim, such as
GRB030329/ SN2003dh) may also have a transrelativistic component.

This component could be identified with the ejection of a thin, fast
shell during the breakout of a shock across the Wolf-Rayet photosphere
(Colgate 1974; Tan, Matzner, \& McKee 2001), which can supply up to
$\sim 10^{48}$ ergs in transrelativistic material; or, in more energetic
events, with a jet cocoon (Ramirez-Ruiz et al. 2002; Pe'er, M\'esz\'aros,
\& Rees 2006).  A third possibility is a shell of Wolf-Rayet material 
that is entrained at the head of a relativistic jet 
(Waxman \& M\'esz\'aros 2003).  Such a `breakout shell' is susceptible 
to fragmentation, and in the case of a wide jet (opening angle tens of degrees)
with a relatively short duration ($t_j < 10$ s), its rest energy can approach
the total jet energy (Thompson 2006).
In this case, the two components will intermix within $\sim 10^{11}-10^{12}$ 
cm from the engine, and may not develop a large Lorentz factor.
(The breakout shell is lighter and accelerates more 
easily in more focused jets with $E_{\rm iso} \ga 10^{52}$ ergs.)

The radius at which gamma-rays are emitted from mildly relativistic ejecta
depends on the ejecta mass and the density of the progenitor wind.  There
are various possibilities: the emission radius could be identified with
the deceleration radius of the relativistic shell (especially if the 
ejecta shell is light and the wind is dense; Tan et al. 2001); with the
photosphere of the wind (Wang et al. 2006); or with the photosphere of the
ejecta themselves.  In the first case, the emission occurs at a time
$t_{\rm em} \sim E_{\rm ej}V_w/2\Gamma_{\rm ej\,0}^4\dot M_w c^3
= 20\,E_{\rm ej\,48}\,V_{w\,8}\,(\Gamma_{\rm ej\,0}/2)^{-4}\,
\dot M_{w\,-4}^{-1}$ seconds in the observer's frame.  (Here 
$V_w$ is the Wolf-Rayet wind velocity, $\dot M_w = \dot M_{w\,-4}\times 10^{-4}
\,M_\odot$ yr$^{-1}$, and $E_{\rm ej}$ is the ejecta energy and 
$\Gamma_{\rm ej\,0}$ is its initial Lorentz factor.)
In the second case, the wind photospheric radius is 
$R_{\tau=1} = Y_e\sigma_T \dot M_w/4\pi V_w m_p$ in the absence of
pair creation, and the emission time is $t_{\rm em} =  
R_{\tau=1}/2\Gamma_{\rm ej}^2c = 
20\,\dot M_{w\,-4}/\Gamma_{\rm ej}^2V_{w\,8}$ seconds (for a wind
composed of alpha elements, $Y_e = {1\over 2}$ electrons per baryon).
In the third case, the ejecta themselves become transparent to scattering
at $t_{\rm em} \simeq (Y_e \sigma_T E_{\rm ej}/16\pi\,\Gamma_{\rm ej}^5\,
m_pc^4)^{1/2} = 2\times 10^3\, E_{\rm ej\,51}^{1/2}\Gamma_{\rm ej}^{-5/2}$
seconds (again neglecting pair creation and assuming 
$Y_e = {1\over 2}$ within the ejecta).  (Note that in this last case, the 
ejecta are too heavy to be decelerated significantly by the Wolf-Rayet
wind before they become optically thin, if the ejecta energy is
as large as $\sim 10^{50}$-$10^{51}$ ergs.)  These timescales can be compared
with the durations of GRB 980425 ($t_{\rm em} \sim 30$ s) and
GRB 060218 ($t_{\rm em}\sim 3\times 10^3$ s for the prompt thermal X-ray component).

Pair creation could evidently be important in the emission of 
GRB 980425, since $E_{\rm pk} \sim 100$ keV was above the threshold for
thermal pair creation ($\sim 30$ keV in the bulk frame).  We consider the
simplest case of pairs in chemical equilibrium with
a Wien gas of photons (density $n_\gamma'$ and mean energy $3k_{\rm B}T'$).
The pair density is related to the photon density by
\be
{n_+' + n_-'\over n_\gamma'} = \left({\pi\over 2}\right)^{1/2}\,
\left({k_{\rm B}T'\over m_ec^2}\right)^{-3/2}\,\exp\left[-{m_ec^2\over 
   k_{\rm B} T'}\right].
\ee
Setting $\tau_T \sim (k_{\rm B}T'/m_ec^2)^{-1}$ (see eq. [\ref{eq:tauTmin}]) 
for the limiting radius of thermalization gives an implicit relation 
for the rest frame temperature,
\be\label{eq:tlrel}
\left({k_{\rm B}T'\over m_ec^2}\right)^{3/2}\,
\exp\left[{m_ec^2\over k_{\rm B} T'}\right] \simeq 0.3\ell',
\ee
where
\be\label{eq:compt}
\ell' = {\sigma_T L_j\over 8\pi \Gamma_{\rm ej}^3m_ec^3 r}
= 6\times 10^2\,E_{j\,48}\left({t_j\over 30~{\rm s}}\right)^{-2}\,
\left({\Gamma_{\rm ej}\over 2}\right)^{-5}
\ee
is the compactness in the emission zone.  When the temperature is
lower than this critical value, it is not possible for the thermal
pairs to upscatter soft keV photons that are advected with the ejecta
from the Wolf-Rayet photosphere.  The energy of the Wien peak
cannot be pushed any lower by sharing the thermal energy
amongst a larger number of photons.   

In the case of GRB 980425, eqs. (\ref{eq:tlrel}) and (\ref{eq:compt}) give 
$k_{\rm B}T' \simeq 50$ keV and $E_{\rm pk} = {4\over 3}\,\Gamma_{\rm ej}
\times 3k_{\rm B}T' = 200\,\Gamma_{\rm ej}$ keV.  
This is about a factor of 3 too hard for $\Gamma_{\rm ej} = 2$, but it
should be recalled that upscattering of soft keV photons freezes out 
at an optical depth $\tau_T \sim 10$.  After the heating rate slows down,
the photons entrained by the optically thick
pair cloud would cool adiabatically.\footnote{The specific heat
of the photons dominates that of the pairs at temperatures well below
$m_ec^2/k_{\rm B}$.  We are interested in the regime where the baryonic
rest energy is comparable to that of the photons, so the baryons
provide inertia.  They do not, however, provide a significant optical depth
if $\ell' \la (m_p/m_e)(\eta/\Gamma)$.  The scattering depth through
the neutralizing electrons can be smaller than unity at the point
where the pairs freeze out, if the pairs are heated continuously starting
at a large optical depth.}
The temperature could drop by by as much as
$\tau_T^{-2/3} = 0.2$ if pairs did not annihilate and
$\Gamma$ remained constant; in fact some deceleration and annihilation 
will take place, which have opposing effects on the output temperature.
A burst duration of $\sim 30$ seconds is obtained if the initial
ejecta Lorentz factor is $\Gamma_{\rm ej\,0} \simeq 2\,\dot M_{w\,-4}^{-1/5}$.  

The X-ray spectrum of GRB 060218 was much too soft for pairs to contribute
significantly to the optical depth.  (The compactness $\ell'$ is inferred
to be only $\sim 10$ in the emission zone if the ejecta are mildly
relativistic.)  The high-energy tail of the X-ray spectrum could be
explained by the Fermi acceleration of soft keV photons 
at the forward shock as it passes through the photosphere of 
the Wolf-Rayet wind, if $\Gamma_{\rm ej} \sim 1-2$  (Wang et al. 2006).
This has the interesting implication that the pre-burst mass loss rate 
is quite high, $\dot M\sim 10^{-2}\, M_\odot$ yr$^{-1}$, in order to place 
the wind photosphere at a distance $c t_{\rm em} \sim 
10^{14}$ cm from the engine (see also Dai, Zhang, \& Liang 2006).   

It also turns out that the radius $ct_{\rm em}$ in GRB 060218
is close to the scattering photosphere of 
the ejecta themselves, if they are mildly relativistic and have a kinetic
energy of $\sim 10^{51}$ ergs.  Therefore a related possibility is that the 
keV photons are warmed up by turbulence within the ejecta (which would be
excited by the differential motion of the relativistic ejecta with respect
to the fragments of the breakout shell; see \S \ref{s:beam}).  It should
be emphasized that the black body radius of $\sim 10^{12}$ cm for the keV
thermal emission (Campana et al. 2006) is much smaller than the photospheric
radius that is inferred from the burst duration.  This is
consistent with the calculation of photon creation in \S \ref{s:therm}.

It is possible that all GRB/SNe combine a transrelativistic outflow with 
a relativistic jet, the relative strengths of the two components 
contributing in varying ratios to the non-thermal $\gamma$-ray emission. 
However, the Amati et al. relation clearly relates to the relativistic 
component, whose presence is inferred in the great majority of 
classical bursts;  whereas in events where the trans-relativistic 
component dominates it appears that the Amati et al. relation is generally 
not satisfied.  Our analysis in \S \ref{sec:23} and \ref{sec:24} applies 
to bursts where the relativistic jet component dominates.

\subsection{Implications for the $E_{\rm pk}$-$E_{\rm iso}$ Relation
of Short GRBs}

The short GRB population offers a nice test of the idea that the
peak energy is fixed by thermalization inside the outflow photosphere.
Although the two burst populations have different progenitors,
the outflow that produces them must have a very high compactness
in both cases, and may be driven by essentially the same mechanism 
(e.g. a MHD jet).  The short GRBs have systematically higher peak energies 
than is implied by the Amati et al. relation (Donaghy et al. 2006),
but so far only GRB 050709 has both a well defined redshift and measured
spectral peak energy (Fox et al. 2005; Villasenor et al. 2005).  

It should be kept in mind that
the black body temperature depends more directly on the outflow
luminosity than the total burst energy.
To obtain the relevant thermalization radius,
we can make use of the scaling (eq. [\ref{eq:Lphot}]),
\be\label{eq:r0scale}
R_0 \propto {\Gamma_0 E_{\rm iso}^{1/2} \over t_j^{1/2} E_{\rm pk}^2}.
\ee
 The isotropic
energy of GRB 050709 is $\simeq 1\times 10^{50}$ ergs and 
its peak energy is $\simeq 100$ keV, some 
10 times higher than would be expected
based on the value of $E_{\rm pk}$ alone.   However, the burst's
$T_{90}$ duration is $\sim 0.07$s, more than 100 times shorter than
a typical long GRB.    Combining equation (\ref{eq:r0scale})
with eq. (\ref{eq:rgamat}) one obtains a thermalization radius of
$R_0 \simeq 1\times 10^{10}$ cm, similar to what we deduced for a long GRB with
$E_{\rm iso} \sim 10^{52}$ ergs s$^{-1}$ and $t_j \sim 10$ s.  

A much smaller thermalization radius is deduced for the giant
flare of 27 December 2004 from SGR 1806$-$20. Here the peak energy
was $\sim 500$ keV and the isotropic luminosity was $4\times 10^{47}$
ergs s$^{-1}$ (Hurley et al. 2005; Palmer et al. 2005).  
The thermalization radius is therefore inferred to be $\sim 50$ km,
about the Alfv\'en radius expected for that luminosity and for
a dipolar magnetic field of $10^{15}$ G.

\section{Implications for the Emission Mechanism}

\subsection{Peak Width-Photon Frequency Correlation}

The light curves of most gamma-ray bursts contain multiple pulses.
A pulse is typically narrower when observed
in a higher-frequency waveband:  one observes the scaling
\be\label{eq:delteg}
\delta t \propto E_\gamma^{-0.5} 
\ee
(Fenimore et al. 1995).

A pulse results from dissipation within some part of
the outflow.  Its duration is limited by particle cooling as well as
by the differential propagation time of the radiation across 
the emitting volume.  We consider both of these effects in turn.

A characteristic pulse width resulting from
dissipation at radius $r$ and Lorentz factor $\Gamma$ is 
$\delta t \sim r/2\Gamma^2 c$.
Multiple pulses of this width would naturally result from dissipation
well inside the radius at which the reverse shock wave passes 
through\footnote{This separation between an inner dissipative zone
where internal shocks (or reconnection events) occur, 
and an outer dissipative zone where 
the afterglow is generated, is an artifact of the assumption that
the gamma-ray emitting jet contains one dominant component.  In fact, 
it is plausible that the outflow contains two components, one of which
is much denser and slower.  The slow component may be derived from
a thin shell of Wolf-Rayet material that is entrained at the jet head 
(Waxman \& M\'esz\'aros 2006).  This breakout shell is susceptible to 
fragmentation, and the shell fragments provide an attractive trigger
for dissipation and gamma-ray emission as they drift backward through
the faster, relativistic jet material (Thompson 2006).  This process
is completed at about $10^{14}$ cm from the engine for $E_{\rm iso}
\sim 10^{52}$ ergs and $t_j \sim 10$ s.}  the ejecta shell 
(e.g. Sari \& Piran 1997).  This broadening effect is geometrical, 
however, and the resulting pulse width is to a first approximation 
independent of frequency.

\subsubsection{Implications for Beaming}\label{s:beam}

Our interpretation of the $E_{\rm pk}$-$E_{\rm iso}$ relation implies that
photons above the spectral peak of a GRB are thermal photons upscattered
in energy.  The existence of a direct mapping between the
spectral peak energy of a GRB and the temperature of the seed thermal photons 
then implies that photons near the spectral peak must be emitted by
material moving in the direction of the observer.  The same conclusion
applies to higher-energy photons.  The frequency dependence of the pulse
profile could also, in principle, depend on the orientation of the
the emitting material with respect to the line of sight (with softer
photons being preferentially emitted off axis).  Our model suggests
that such orientation effects are of secondary importance in 
explaining the correlation (\ref{eq:delteg}).

This observation has important consequences for the emission mechanism.
The scaling (\ref{eq:delteg}) is suggestive of the cooling of 
relativistic particles:  the characteristic frequency 
of the synchrotron or inverse-Compton photons emitted by an electron 
of energy $\gamma_e\,m_ec^2$ is proportional to $\gamma_e^2$, and the cooling 
time is inversely proportional to $\gamma_e$.   The pulse width is controlled
by cooling (over some range of frequencies) only if the light travel time
across the radiating plasma is shorter than the cooling time
at each (observed) 
frequency.   This implies that the size of the cooling region is
$L'/ct' \la (\ell\gamma_e)^{-1}$, where $\ell \ga 1$ is the compactness
in the bulk frame.

One can consider dividing up each causally connected patch in the outflow
(of size $ct' \sim r/2\Gamma$) into cells of size $\sim ct'/\gamma_e$.  A
single 
causal patch then comprises $\sim \gamma_e^3$ such cells.  If the emission 
is nearly isotropic in each cell, then the emission from any of them will 
be detectable.  On the other hand, if the emission is beamed into a solid
angle $\sim 1/\gamma_e^2$, then only $\sim \gamma_e$ cells are visible 
to any observer.  In this case, the pulse duty cycle is given simply by
the fraction of cells that experience strong dissipation.  

Strong beaming
of this type is not expected if the cooling particles are accelerated
at a shock.   Two mechanisms for beaming have been suggested in the
case where the outflow is magnetically dominated:  bulk relativistic motion
of the magnetofluid (due to, e.g., current-driven instabilities;
Lyutikov \& Blandford 2003); and heating of the motion of the light 
charges parallel to the background magnetic field, due to Landau damping of 
high-frequency Alfv\'enic turbulence (Thompson \& Blaes 1998; Thompson 2006). 

An additional form of beaming is expected if the outflow carries
a slower and denser component.  The fragmentation of the heavy component
occurs over a range of angular scales as it is accelerated outward by
the momentum flux of the lighter relativistic fluid.  At a distance
of $\sim 10^{14}$ cm from the engine, the differential Lorentz factor
between the two components has been reduced to $\Delta\Gamma \sim 2$
(Thompson 2006).  A fraction of the seed thermal radiation will side-scatter
off the heavy component (which remains optically thick out to 
$\sim 10^{14}$ cm from the engine), and can provide an enhanced coolant
for relativistic particles in the relativistic fluid. (See
Sikora, Begelman \& Rees 1994 for related considerations in the context of
Blazar jets.)  The inverse Compton radiation so produced will be preferentially
beamed with respect to the side-scattered thermal radiation
by a modest Doppler factor $\sim \Delta\Gamma$.

\subsubsection{Inverse Compton Cooling and Other Cooling Mechanisms}

A simple relation
between cooling time and photon frequency is expected if each high-energy photon results
from a single upscattering.  The cooling time in the bulk frame of the outflow is 
$t_{\rm cool}' = 3m_ec/4\gamma_e \sigma_T U_\gamma'$, and the observed cooling
time $t_{\rm cool}$ is shorter by a factor $1/2\Gamma$.  Setting
\be
E_\gamma = \gamma_e^2 E_{\rm pk}
\ee
and making use of the Amati et al. relation (eq. [\ref{eq:amati}]), gives
\be\label{eq:tcoolic}
t_{\rm cool}({\rm IC}) = 0.8\;{r_{15}^2 \Gamma_2 t_{j\,1}\over E_{\rm iso\,52}^{3/4}}\,
\left({E_\gamma\over 100~{\rm keV}}\right)^{-1/2}\;\;\;\;{\rm s}.
\ee
(Here $E_{\rm iso}$ is the isotropic energy of the prompt emission.)
Pulses of a width of $\sim 1$ s must be emitted within $\sim 10^{15}$ cm
from the engine, if the outflow Lorentz factor is close to the limiting value
of $\sim 10^2$ (eq. [\ref{eq:gamax}]).   Requiring that the cooling
time (\ref{eq:tcoolic}) be shorter than $r/2\Gamma^2c$ implies
\be\label{eq:rmax}
r < 2\times 10^{15}~{E_{\rm iso\,52}^{3/4}\over \Gamma_2^3\, t_{j\,1}}
\,\left({E_\gamma\over 100~{\rm keV}}\right)^{1/2}\;\;\;\;{\rm cm}.
\ee

Pair creation in the GRB outflow just outside the baryonic photosphere has a strong
influence on the pulse duration.  The prefactor $\Gamma_2^{-3}$ could be
as small as $\sim 10^{-2}$ if the outflow became heavily pair loaded
while being accelerated by the anisotropic pressure of the fireball photons
(eq. [\ref{eq:gamax}]).  Substituting the upper bound (\ref{eq:rmax})
on $r$ back into equation (\ref{eq:tcoolic}) for $t_{\rm cool}({\rm IC})$,
one deduces
\be\label{eq:tcoolicb}
t_{\rm cool}({\rm IC}) < 0.01\,{E_{\rm iso\,52}^{3/4}
\over (\Gamma_2/3)^5\,t_{j\,1}}
\left({E_\gamma\over 100~{\rm keV}}\right)^{1/2}\;\;\;\;{\rm s}.
\ee
Broader pulses can, of course, result from dissipation in the outflow
outside the radius (\ref{eq:rmax}), but their duration is limited
by causality and not by IC cooling. 

A more energetic
particle is required to emit a synchrotron photon of a given energy, 
which means that the synchrotron cooling time is correspondingly shorter.
The equivalent result for synchrotron emission may be obtained by
setting $B'^2/8\pi = (\varepsilon_B/\varepsilon_\gamma)U_\gamma'$ and
$E_\gamma' = 0.3\gamma_e^2 \hbar eB'/m_ec$.  One finds
\be
{t_{\rm cool}({\rm synch})\over t_{\rm cool}({\rm IC})} = 
2\times 10^{-3}\,{\varepsilon_\gamma^{7/8}\over
\varepsilon_B^{3/4}}\cdot{L_{\rm iso\,51}^{1/8}\over r_{15}^{1/4}\Gamma_2^{1/4}}.
\ee
Pulses as broad as $\sim 1$ s, with the frequency-dependent width
observed in GRB light curves, cannot easily be explained by
synchrotron emission.

A frequency-dependent pulse width would also result from the damping
of bulk relativistic motion by Compton drag, but in that case the
scaling with photon frequency would be different.  Suppose that a small blob
of speed $\Delta\beta_0$ 
[Lorentz factor $\Delta\Gamma_0 = (1-\Delta\beta_0^2)^{-1/2}$]  is created in
the outflow (e.g. by relativistic reconnection).   As it slows down
to a speed $\Delta\beta < \Delta\beta_0$, the size of the blob increases 
to $\sim (1-\Delta\beta)ct_{\rm drag}'$, where $t_{\rm drag}'$ is the drag
time at Lorentz factor $\Delta\Gamma$.
The duration of the emission is therefore $\delta t \propto
t'/\Delta\Gamma^3$.  Since the energy of the inverse Compton (IC) photons is
$E_\gamma \propto \Delta\Gamma^2$, one finds 
the scaling $\delta t \propto E_\gamma^{-3/2}$, much stronger than is observed.

The frequency dependence of the pulse width at high and low frequencies
deserves special scrutiny, for two reasons.
\begin{enumerate}
\item
The size of the emitting cells in the outflow is more or less independent of the
maximum Lorentz factor $\gamma_e$ of the heated particles.  This means
that observations of gamma-ray pulses at energies well above $\sim 1$ MeV
may provide a direct diagnostic of the cell size:  above some characteristic
value of $E_\gamma$, the observed pulse width should saturate at some minimum
value $\sim L'/c\Gamma$. 
\item
Photons detected below the spectral peak have not been upscattered by
particles with relativistic energies in the bulk frame.  The pulse width 
should, as a result, have a weaker frequency dependence below the spectral
peak.
\end{enumerate}

We can summarize our conclusions as follows.  First, the
frequency-dependence of the pulse width arises most naturally from
IC cooling of {\it random} particle motion.  Since the cooling timescale is
much shorter than the flow time in the emitting frame, the dissipation
producing each pulse must be localized at some radius.  This implies
that the motion of the heated particles is strongly anisotropic; otherwise the 
emission on shorter timescales (at higher frequencies) would be smeared 
out by the differential propagation delay associated with shell curvature.

\subsection{Implications for the Mechanism of Particle Heating}

The temperature of the seed thermal photons is independent of 
radius (in the absence of adiabatic cooling).  
The peak energy resulting from IC scattering of the thermal photons
therefore depends most directly on the mechanism by which the particles are 
heated:  it is sensitive to the
distribution of particle energies.  When the 
heated particles have a power-law energy distribution that is cut
off from below  at $\gamma_{\rm min}\,m_ec^2$, it is clear that
$E_{\rm pk}$ will remain close to the thermal peak energy 
only if $\gamma_{\rm min} \sim 1$.

A value of $\gamma_{\rm min}$ close to unity is easily achieved
if the particles are heated continuously, and the outflow is
at least moderately optically thick to scattering.
Stochastic acceleration by turbulence (Dermer, Miller, \& Li 1996; 
Thompson 2006) or by electrostatic acceleration in reconnection
layers (Romanova \& Lovelace
1992; Larrabee, Lovelace \& Romanova 2003) are possible heating
mechanisms.  This implies that the high-energy tail of the prompt
GRB spectrum is generated near the scattering photosphere.

The synchrotron self-Compton process (Ghisellini \& Celotti 1999;
Stern \& Poutanen 2004) requires $\gamma_{\rm min} \gg 1$.  If 
the particle heating is truly continuous, then this condition can
only be satisfied outside the scattering photosphere.  In that
case, $\gamma_{\rm min}^2 \sim \tau_T^{-1} \propto r \Gamma^2\eta/L_{\rm iso}$.
The self-absorption frequency is given
by $h\nu_{\rm sa} \sim \Gamma\,\gamma_{\rm sa}^2 (B'/B_{\rm QED})m_ec^2
\propto L_{\rm iso}^{5/14}\,\Gamma^{2/7}\,r^{-5/7}$, where $\gamma_{\rm sa}$
is given by eq. (\ref{eq:gamsatb}).  The peak of the 
Comptonized synchrotron spectrum is at
\be
E_{\rm pk} \sim \gamma_{\rm min}^2\,h\nu_{\rm sa} \propto 
\eta \Gamma^{16/7}\,L_{\rm iso}^{-9/14}\,r^{2/7}.
\ee
The explicit dependence on radius nearly scales out here,
but the exponent of $L_{\rm iso}$ has the wrong sign.
(In this model, it is not clear how $\Gamma$ and $L_{\rm iso}$ are related,
since the seed thermal radiation field must be assumed to be absent.)

The continuous heating of
seed thermal photons by a second-order Compton process has also
been considered (Thompson 1994; Giannios 2006); but this mechanism does not
provide an obvious explanation for the correlation 
(\ref{eq:delteg}) between pulse width and photon frequency, 
or the lags of soft photons with respect to hard photons that are 
observed in some GRBs (especially those with broad pulses, e.g. 
Norris et al. 2005).

\subsubsection{Shock Acceleration and $\gamma_{\rm min} \sim 1$}

Shock acceleration is a rapid process:  particles reach an
energy $\gamma_e m_ec^2$ on a timescale comparable to the 
gyroperiod $\gamma_e m_e c/eB$.  Only modest
Lorentz factors ($\gamma_e\sim1$-10) are needed to create an IC spectrum 
extending up to $\sim 10^2\,E_{\rm pk}$.  At such low energies, the 
particle distribution is modified by cooling only at a large distance
downstream of the shock.

Under what conditions can shock acceleration generate a particle
distribution with $\gamma_{\rm min} \sim 1$, with a significant
fraction of the outflow energy deposited in the non-thermal particles?   
Three basic requirements must be satisfied:
first, a large fraction of the particle inertia 
in the outflow must be in light charges (electrons and positrons);
second, a significant fraction of the light charges which encounter
a shock must undergo Fermi acceleration; 
and, third, the momentum distribution of the of accelerated particles must
have an index close to $-2$ (or harder).   Although
the behavior of a pair-dominated, relativistic shock 
is not fully understood (see Hoshino et al.
1992 for particle-in-cell simulations of electron-positron-ion
shocks), these considerations are largely independent of such details.

Pair creation in the outflow occurs primarily through collisions
between photons (Cavallo \& Rees 1978; Baring \& Harding 1997;
Lithwick \& Sari 2001).   These pair-creating photons have
a characteristic energy $\sim m_ec^2$ in the bulk frame,
and are a byproduct of the cooling of charges of an energy
$\gamma_e \sim (\Gamma\,m_ec^2/E_{\rm pk})^{1/2}$.  Cooling is
faster than pair creation, and so the equilibrium pair density
can be determined by balancing the rates of pair creation and
annihilation.  Essentially the entire energy of the injected non-thermal
particles is converted to a non-thermal photon spectrum by rapid
IC cooling.

To simplify matters, 
we assume that this process occurs over some range of radius in
the outflow, so that the pair density is already close to its equilibrium
value.  For a high-energy photon index $\beta \geq 2$, 
the positron creation rate is $\dot n_+' 
\simeq (0.1\,\sigma_T)n_\gamma'(m_ec^2)$
(Svensson 1987), where $n_\gamma'(m_ec^2)$ is the photon density 
at the threshold energy $E_\gamma' = m_ec^2$.   Balancing this with the
annihilation rate $\dot n_+' = -(3/8)\sigma_T c(n_+')^2$ in a warm pair plasma
gives
\be\label{eq:npmeq}
{(n_+' + n_-')m_ec^2\over U_\gamma'} \simeq \varepsilon_{\rm nth}\,(\beta-2)\,
\left({E_{\rm pk}'\over m_ec^2}\right)^{\beta-2} \simeq 
\varepsilon_{\rm nth}(\beta-2)\,
\left({E_{\rm pk}\over \Gamma m_ec^2}\right)^{\beta-2}
\ee
for rest energy density in the created pairs.  The total photon energy density
$U'_\gamma$ can be divided into a non-thermal tail carrying a fraction
$\varepsilon_{\rm nth}$ of the total, and a thermal peak that has
been boosted in energy by the cooling of the remaining thermal pairs.  
Now $\gamma_{\rm min}$ is minimized if a significant fraction 
of the pairs are converted to a non-thermal 
distribution at the shock. The net energy density deposited in 
non-thermal pairs with particle index $p$ is then
\be
{\gamma_{\rm min}\over (p-2)}m_ec^2 (n_+'+n_-') \simeq \varepsilon_{\rm nth}
U_\gamma'.
\ee
Combining this expression with (\ref{eq:npmeq}) and the relation
$p-2 = 2(\beta-2)$ for rapidly cooling particles gives
\be
\gamma_{\rm min} \simeq 2\,\left({E_{\rm pk}\over \Gamma m_ec^2}\right)^{-(\beta-2)}.
\ee
For example, if $\beta = 2.2$, $E_{\rm pk} = 100$ keV  and $\Gamma = 
10^2$ then $\gamma_{\rm min} = 7$.  Even a modest departure of the
photon spectrum from a $-2$ index implies that $\gamma_{\rm min}$ is
large enough to give a considerable IC boost
to the energy of the seed thermal photons.

Softer high-energy spectra are observed in many GRBs (e.g. Preece et al.
2000).  These spectra could still be consistent with a hard IC spectrum 
in the emission zone, 
if the high-energy photon flux were degraded by pair creation at a larger
radius (e.g., in a dense Wolf-Rayet wind: Thompson \& Madau 2000; 
M\'esz\'aros, Ramirez-ruiz, \& Rees 2001; Beloborodov 2002).

\section{Conclusions}

Most models of GRB outflows assume that they contain a single dominant
component:  e.g. baryons with a variable Lorentz factor (in the case of
internal shock models)  or a non-radial magnetic field with a reversing
sign (in the case of reconnection models).   Our central argument in
this paper is that a second component is essential for understanding the
prompt emission of GRBs:  blackbody radiation that is emitted where
the GRB jet forces its way through the core of a Wolf-Rayet star.  
Its thermal peak (Doppler boosted by the outflow) is identified with 
$E_{\rm pk}$. 
The non thermal high energy part of the GRB emission arises from 
Comptonization of this radiation by relativistic electrons outside the 
effective photosphere. 

This model accounts naturally for the small scatter 
in the Amati et al. and Firmani et al. relations.  
It should be re-iterated that both these relations
can be tied directly to the jet properties in the thermalization 
zone, and are reproduced without free parameters by a very
simple model.  (Indeed, the relative dependence of $E_{\rm pk}$
on the jet energy and duration that was measured by Firmani et al. 2006 
was anticipated in the analysis of this model in Thompson 2006.)
If we assume that the radius of prompt thermalization
is fixed at the progenitor core radius (or some other characteristic 
radius), then it is possible to
reproduce the Ghirlanda et al. relation for a particular dependence
of jet energy on opening angle, but not one that is motivated by
a simple argument.  A key, unresolved
question centers on how the relativistic jet material builds up in the core
of a Wolf-Rayet star:  how wide is the outflow from the central engine,
and how tightly regulated is the total energy that is deposited in the
relativistic fluid with respect to the core binding energy?  

This quasi-thermal model for the observed spectral correlations has
interesting implications for the limiting jet Lorentz factor, 
and the relation of the jet  energy, opening angle
and burst duration to the mass and radius of the stellar stellar progenitor.
The observed relation between pulse width and photon frequency can be
explained by Compton cooling, but one requires that the relativistic particles
in the outflow have an energy distribution with a low-energy 
(trans-relativistic) cutoff, and that the IC emission is beamed.  
The relation between $E_{\rm pk}$ and $E_{\rm iso}$ becomes more complicated
for bursts with isotropic energies less than $\sim 10^{52}$ ergs:  
the spectrum can be harder than is implied by the Amati et al. 
relation if the jet is clean but covers a large solid angle, or if the 
ejecta mass and energy are low.

\acknowledgments
The authors would like to thank Asaf Pe'er for pointing out a few
typographical errors in the equations.  
We acknowledge the support of the NSERC of Canada (CT), NASA NAG5-13286 and
NSF AST 0307376 (PM) and PPARC (MJR).

{}

\end{document}